\documentclass[twocolumn,prx,aps,superscriptaddress,longbibliography]{revtex4-2}
\usepackage{amsmath}
\usepackage{amssymb}
\usepackage{amsfonts}
\usepackage[dvips]{graphicx}
\usepackage{subfigure}
\usepackage{dcolumn}
\usepackage{txfonts}
\usepackage{bm}
\usepackage{makeidx}
\usepackage{color}
\usepackage{mathtools}
\usepackage{threeparttable}
\usepackage[colorlinks,linkcolor=blue,anchorcolor=blue,citecolor=blue,urlcolor=blue]{hyperref}

\begin{document}


\title{Geometric phase for twisted light}
\author{Li-Ping Yang}
\email{lipingyang87@gmail.com}
\affiliation{Center for Quantum Sciences and School of Physics, Northeast Normal University, Changchun 130024, China}

\begin{abstract}
Polarization vectors of light traveling in a coiled optical fiber rotate around its propagating axis even in the absence of birefringence. This rotation was usually explained due to the Pancharatnam-Berry phase of spin-1 photons. Here, we use a purely geometric method to understand this rotation. We show that similar geometric rotations also exist for twisted light carrying orbital angular momentum (OAM). The corresponding geometric phase can be applied in photonic OAM-state-based quantum computation and quantum sensing. 
\end{abstract}

\maketitle

\textit{Introduction}---In the 1980s, the geometric rotation of polarization vectors of light was observed in a helically wound single-mode fiber by Ross~\cite{ross1984rotation} and other researchers~\cite{varnham1985helical}. This phenomenon was later explained due to the Pancharatnam-Berry phase~\cite{pancharatnam1956generalized,berry1984quantal} $\gamma(\mathcal{C})=-2\pi m_s (1-\cos\theta)$ of spin-1 photons ($m_s=\pm1$)  travelling on a helix with pitch angle $\theta$ by Chiao and Wu~\cite{Chiao1986manifestations}. Tomita and Chiao experimentally verified this photonic Pancharatnam-Berry phase of more general fiber configurations with non-uniform torsion~\cite{Tomita1986observation}. The photonic geometric phase inspired streams of research in polarization-dependent Hall effect of light~\cite{Onoda2004Hall,hosten2008observation,Bliokh2008,yin2013photonic,bliokh2015spin,cohen2019geometric}. 

In Chiao and Wu's quantum description~\cite{Chiao1986manifestations}, the helicity of photons $\hat{\boldsymbol{S}}\cdot \boldsymbol{k}/|\boldsymbol{k}|$, which is the projection of photon spin on the wave vector $\boldsymbol{k}$ axis, is an adiabatic invariant during propagation. Thus, eigenstates of the photon spin could accumulate a Pancharatnam-Berry phase when moving in the parameter (reciprocal) space. Berry constructed  a Schr{\"o}dinger-like equation with an effective photon-spin Hamiltonian for electromagnetic fields propagating in a mono-mode fiber to complete this quantum interpretation~\cite{berry1990quantum}. In addition to the photon helicity, the projection of the photonic OAM in the propagating direction is also conserved~\cite{allen1992Orbital,Yang2021nonclassical}. An interesting question arises does a similar geometric rotation exist for the OAM degrees of freedom of light~\cite{van1993geometric}? In this work, we show the answer is yes. 

In parallel to the quantum Pancharatnam-Berry phase description~\cite{Chiao1986manifestations,berry1990quantum}, classical geometric interpretations of the anholonomy of coiled light have also been proposed~\cite{ross1984rotation,berry1987interpreting,haldane1986path}, including the pioneering work by Vladimirski{\u{\i}}~\cite{vinitskiui1990topological}. Inspired by these insights, we give a different explanation for this rotation by combining the differential geometry of the coiled fiber path and reflections of light in the fiber. The key idea is to evaluate the rotation of the photon coordinate frame (PCF) with respect to the local coordinate frame (LCF) parallelly transporting on the fiber. We prove that the LCF always recovers its initial configuration after one circulation if the two ends of the fiber are parallel. The de-synchronization between the PCF and LCF leads to the rotation of the light polarization vectors, i.e., the previous geometric phase of photon spin states~\cite{Chiao1986manifestations}. Applying our theory to a twisted light, we show that light carrying $m \hbar$  OAM will acquire a geometric phase $\gamma_m(\mathcal{C})=2m\pi\cos\theta$ after one helical circulation.

\begin{figure}
\centering
\includegraphics[width=10cm]{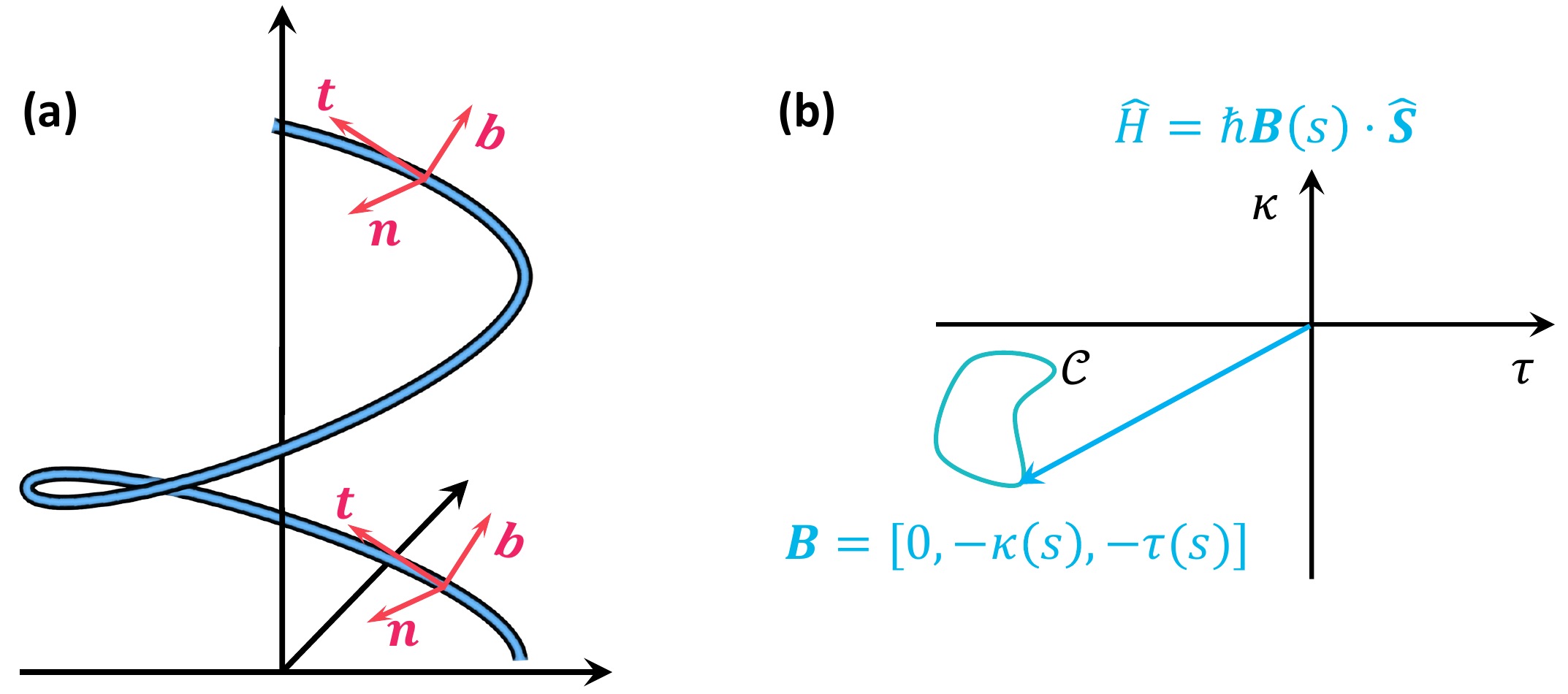}\caption{\label{fig:1} (a) Parallel transport of the local coordinate frame (LCF) on a curve. (b) Movement of the LCF in the parameter space described by the effective magnetic field $\boldsymbol{B}(s)=[0,-\kappa (s),\tau (s)]$ and Hamiltonian $\hat{H}=\hbar\boldsymbol{B}\cdot\hat{\boldsymbol{S}}$. Here, $\kappa (s)$ and $\tau (s)$ are the local curvature and torsion of the curve at $s$, and $\hat{\boldsymbol{S}}$ is an operator vector composed of generators of the SO(3) rotations as given in Eq.~(\ref{eq:generators}). }
\end{figure}

We recently found that our geometric phase for twisted light is the same phase given by Alexeyev and Yavorsky~\cite{Alexeyev2007,alexeyev2006topological}. By re-expressing Maxwell equations in the LCF via a coordinate transformation, Alexeyev and Yavorsky showed that the non-vanishing torsion will lead to an effective twisting for the PCF. All vectors co-moving with light will be rotated in the same way shown in the following. Our work provides an alternative way to understand this rotation. This geometric rotation is universal for waves propagating along a coiled path, such as acoustic waves~\cite{Wang2018topological} and matter waves. A similar effect could also be observed for an electron or atom beam traveling along a coiled ideal-reflection tube.

\textit{Parallel transport of the local coordinate frame}---The unit tangent vector $\boldsymbol{t}(s)$, normal vector $\boldsymbol{n}(s)$,
and binormal vector $\boldsymbol{b}(s)$ form an LCF, which will be parallelly transported along the fiber as shown in Fig.~\ref{fig:1} (a). When a curve is parameterized by its arc length $s$, the dynamics of the LCF are governed by Frenet–Serret formulas~\cite{do2016differential}
\begin{align}
\frac{d}{ds}\boldsymbol{n}(s) & =-\kappa(s)\boldsymbol{t}(s)+\tau(s)\boldsymbol{b}(s),\\
\frac{d}{ds}\boldsymbol{b}(s) & =-\tau(s)\boldsymbol{n}(s),\\
\frac{d}{ds}\boldsymbol{t}(s) & =\kappa(s)\boldsymbol{n}(s),
\end{align}
where $\kappa(s)$ and $\tau(s)$ are the curvature and torsion of
the curve at $s$, respectively. In the following, we show that the photonic geometric rotation cannot be explained simply by the parallel transport of the LCF. 

For a uniform helix, its curvature $\kappa$ and torsion $\tau$ are constants. A helix with radius $a$ and pitch $2\pi b$ is usually described by the parametric vector in the stationary laboratory reference frame [denoted by $xyz$ coordinates in Fig.~\ref{fig:1} (a) ]~\cite{do2016differential},
\begin{equation}
\boldsymbol{r} (s) = \left(a\cos\frac{s}{c}, a\sin\frac{s}{c}, b\frac{s}{c} \right), 
\end{equation}
where $c^2 = a^2 +b^2$ and the parameters are connected to the curvature and torsion via
\begin{equation}
\kappa = \frac{a}{c^2},\ \tau = \frac{b}{c^2}.    
\end{equation}
We can re-parameterize the helical curve with the azimuthal angle $\varphi = s/c = \omega s$ with $\omega=\sqrt{\kappa^{2}+\tau^{2}}=1/c$. We note that $\omega$ is the spatial  frequency with respect to the azimuthal angle $\varphi$ and it is not the frequency of light. The uniform helix is now described by
\begin{equation}
\boldsymbol{r} (\varphi) =\frac{1}{\omega^2} \left(\kappa\cos\varphi, \kappa\sin\varphi, \tau\varphi \right). 
\end{equation}
One round trip $\mathcal{C}$ of light in Ref.~\cite{Chiao1986manifestations} corresponds to our azimuthal angle $\varphi$ varying from $0$ to $2\pi$.

The dynamic evolution of the LCF on a uniform helix is given by $[\boldsymbol{n}(\varphi), \boldsymbol{b}(\varphi), \boldsymbol{t}(\varphi)]^T = \hat{R}(\varphi,\boldsymbol{e})[\boldsymbol{n}(0), \boldsymbol{b}(0), \boldsymbol{t}(0)]^T$. Here,
\begin{align}
\hat{R}(\varphi,\boldsymbol{e}) = \left[\begin{array}{ccc}
\cos\varphi & -\cos\theta\sin\varphi & \sin\theta\sin\varphi\\
\cos\theta\sin\varphi & \sin^{2}\theta+\cos^{2}\theta\cos\varphi & \sin2\theta\sin^2\frac{\varphi}{2}\\
-\sin\theta\sin\varphi & \sin2\theta\sin^2\frac{\varphi}{2} & \sin^{2}\theta\cos\varphi+\cos^{2}\theta
\end{array}\right],\label{eq:R}
\end{align}
is the rotation operator of a vector in the three-dimensional Euclidean space around the axis $\boldsymbol{e}=[0,\sin\theta,\cos\theta]$ with $\sin\theta=\kappa/\omega$ and $\cos\theta=\tau/\omega$. We note that our definition of the angle $\theta$ is the same as the pitch angle used in Tomita and Chiao's experiment~\cite{Tomita1986observation}, but different from Rytov's rotation angle~\cite{kravtsov1990geometrical,vinitskiui1990topological}. Previously, the Frenet equations were solved via the mapping between SO(3) and the SU(2) Lie algebras~\cite{berry1993classical}. 

We emphasize that the geometric rotation of the polarization vector of light traveling in a uniform helical fiber cannot be explained with the parallel transport of the LCF. We can verify that the rotation operator $\hat{R}(\varphi,\boldsymbol{e})$ equals the identity matrix $\hat{I}$ for $\varphi=2\pi$. Thus, no rotation of the LCF occurs after an adiabatic evolution loop. The 
LCF always returns its initial configuration in the $xyz$-frame. This can also be seen from the explicit expressions of the LCF unit vectors in the $xyz$-frame
\begin{align}
\boldsymbol{t}(s)&=\frac{d\boldsymbol{r}(s)}{ds}=(-\sin\theta\sin\varphi,\sin\theta\cos\varphi,\cos\theta), \label{eq:t_s}\\
\boldsymbol{n}(s)&=\frac{1}{\kappa}\frac{d^{2}\boldsymbol{r}(s)}{d^{2}s}=(-\cos\varphi,-\sin\varphi,0),\label{eq:n_s}\\  \boldsymbol{b}(s)&=\boldsymbol{t}(s)\times\boldsymbol{n}(s)=(\cos\theta\sin\varphi,-\cos\theta\cos\varphi,\sin\theta).\label{eq:b_s}
\end{align}
We see that these three unit vectors are the same for $\varphi = 0$ and $\varphi =2\pi$. 
Next, we will show that this result is still valid for a non-uniform helix if two ends of the curve are parallel. 


The dynamics of the LCF on an arbitrary singular-free curve are governed by the Frenet-Serret formulas, which can be re-expressed as a Schr{\"o}dinger-like equation
\begin{equation}
i\frac{d}{ds}\left|\psi(s)\right\rangle =\hat{H}(s)\left|\psi(s)\right\rangle, \label{eq:Schrodinger}
\end{equation}
where $\left|\psi(s)\right\rangle =\left[\boldsymbol{n}(s),\boldsymbol{b}(s),\boldsymbol{t}(s)\right]^{T}$
and the Hamiltonian is given by
\begin{equation}
 \hat{H}(s)=\boldsymbol{B}(s)\cdot\hat{\boldsymbol{S}},  \label{eq:H} 
\end{equation}
with an effective magnetic field $\boldsymbol{B}(s)=[0,-\kappa(s),-\tau(s)]$
and the generators of the SO(3) rotations
\begin{equation}
\hat{S}_{n}=\left[\begin{array}{ccc}
0 & 0 & 0\\
0 & 0 & -i\\
0 & i & 0
\end{array}\right],\ \hat{S}_{b}=\left[\begin{array}{ccc}
0 & 0 & i\\
0 & 0 & 0\\
-i & 0 & 0
\end{array}\right],\ \hat{S}_{t}=\left[\begin{array}{ccc}
0 & -i & 0\\
i & 0 & 0\\
0 & 0 & 0
\end{array}\right].\label{eq:generators}
\end{equation}
By diagonalizing Hamiltonian (\ref{eq:H}), we obtain the instantaneous eigenstates of $\hat{H}(s)$ 
\begin{align}
\left|+1\right\rangle  & =\frac{1}{\sqrt{2}}\left[-i\boldsymbol{n}(s),-\cos\theta (s)\boldsymbol{b}(s),\sin\theta (s)\boldsymbol{t}(s)\right]^{T},\\
\left|0\right\rangle  & =\left[0,\sin\theta (s)\boldsymbol{b}(s),\cos\theta (s)\boldsymbol{t}(s)\right]^{T},\\
\left|-1\right\rangle  & =\frac{1}{\sqrt{2}}\left[i\boldsymbol{n}(s),-\cos\theta (s)\boldsymbol{b}(s),\sin\theta (s)\boldsymbol{t}(s)\right]^{T},
\end{align}
with corresponding eigenvalues $0$ and $\pm \omega (s)$. We note that the curvature $\kappa (s)$, torsion $\tau (s)$, $\omega(s)=\sqrt{\kappa^{2}(s)+\tau^{2}(s)}$, and the pitch angle  $\theta(s)$ are all dependent on $s$ for a non-uniform helix.

Now, we verify that the unit vectors of the LCF always return to their initial configuration after an adiabatic circulation.  Round a close curve $\mathcal{C}$ in the $\boldsymbol{B}$-parameter space [see Fig.~\ref{fig:1} (b)], the eigenstate $|n\rangle$ ($n=0,\ \pm 1$) of $\hat{H}$ will accumulate both the dynamic phase $\alpha_{n}(\mathcal{C})$
and the Pancharatnam-Berry phase $\gamma_{n}(\mathcal{C})$ from perspective of the effective Schrodinger equation (\ref{eq:Schrodinger})~\cite{pancharatnam1956generalized,berry1984quantal}. We note that the adiabatic parameter is the effective magnetic field $\boldsymbol{B}(s)$ not the wave vector as in Wu and Chiao's work~\cite{Chiao1986manifestations}. For a non-uniform helix, both the curvature and the torsion will not change signs, thus the point $\boldsymbol{B}=[0,0,0]$ will not be enclosed by the path $\mathcal{C}$  in the parameter space. The Pancharatnam-Berry phases for all three
states are zero since the corresponding solid angle $\Omega_{n}(\mathcal{C})$ in the parameter space vanishes as shown in Fig.~\ref{fig:1} (b). The dynamics phases for the three eigenstates are given by
\begin{equation}
\alpha_{n}(\mathcal{C})=-n\int_{\mathcal{C}}\omega(s)ds=-n\int_{0}^{2\pi}d\varphi=-2n\pi,\ n=0,\pm1.
\end{equation}
Here for each small segment, we have used the geometric relation $\omega (s) ds=d\varphi$, which is different from the one $d\varphi/ds=\tau$ in Rytov's rotation~\cite{vinitskiui1990topological}. We note that the dynamic phase $\alpha_n(\mathcal{C})$ originates from the variation of the LCF on an arbitrary curve and is not the propagating phase of light. The propagating phases, which are determined by the length and the permittivity of the fiber, are of the same non-zero value for both a straight fiber and a coiled fiber. However, $\alpha_n(\mathcal{C})$ is always zero for a straight fiber.  By expanding $\boldsymbol{n}(s)$, $\boldsymbol{b}(s)$, and $\boldsymbol{t}(s)$ with \{$|n\rangle$\}, we can verify that the LCF, as well as an arbitrary vector $\boldsymbol{v}$ co-moving with the LCF, returns to its initial configuration in the $xyz$-frame round $\mathcal{C}$. This proves our claim that the LCF remains the same after an adiabatic circulation.

\begin{figure}
\centering
\includegraphics[width=8.5cm]{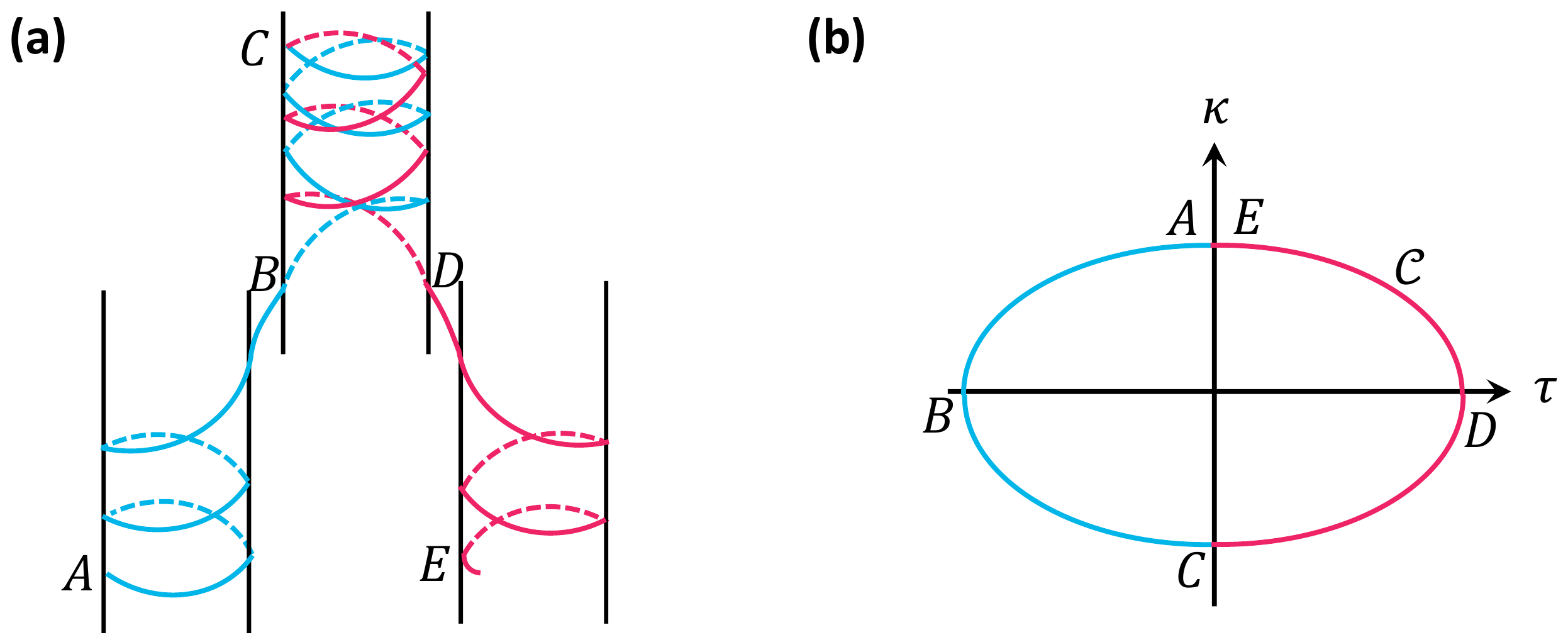}\caption{\label{fig:2} (a) A winding method of fiber to obtain non-vanishing Pancharatnam-Berry phase for the local coordinate frame (LCF). (b) The path of the LCF in the parameter space. At points $B$ and $D$, the curvature $\kappa (s)$ of the curve changes sign. At point $C$, the sign of the torsion $\tau (s)$ of the curve flips.}
\end{figure}

Different from a M\"obius strip, a helix is topologically trivial. Thus, no twisting of the LCF will occur during the parallel transport as expected. We emphasize that the vector $\boldsymbol{n}(s)$, $\boldsymbol{b}(s)$, or the circular polarization vector  $\boldsymbol{e}_{\pm}\equiv (\boldsymbol{n}\pm i\boldsymbol{b})/\sqrt{2}$ is not an eigenstate of $\hat{H}(s)$. Thus, an Aharonov-Anandan (AA) phase~\cite{Aharonov1987phase} instead of a Pancharatnam-Berry phase will be obtained via
\begin{equation}
\gamma_{AA}=  i\int_{\mathcal{C}} ds \langle\psi(s)|\partial_s\psi(s)\rangle = i\int_0^{2\pi} d\varphi \langle\psi(\varphi)|\partial_\varphi\psi(\varphi)\rangle,
\end{equation}
where $|\psi(s)\rangle$ is given by the corresponding expansion coefficients in the basis $\{|n(s)\rangle\}$.
The AA phase can not be used to explain the geometric rotation of the polarization vector~\cite{ross1984rotation}, because the no observable phase difference between $|\psi (0)\rangle$ and $|\psi (\varphi = 2\pi )\rangle$ exists as explained above. In principle, we can construct a complicated winding path as shown in Fig~\ref{fig:2} (a), such that a non-vanishing Pancharatnam-Berry for the LCF can be obtained. Special care needs to be taken at points $B$ and $D$ in Fig~\ref{fig:2} (b), at which the curvature $\kappa (s)$ vanishes and the normal vector $\boldsymbol{n}$ is not defined. More details about the non-vanishing Pancharatnam-Berry phase for the LCF are out of our interest in this work. Next, we focus on the rotation of the PCF with respect to the LCF.  

\begin{figure}
\centering
\includegraphics[width=10cm]{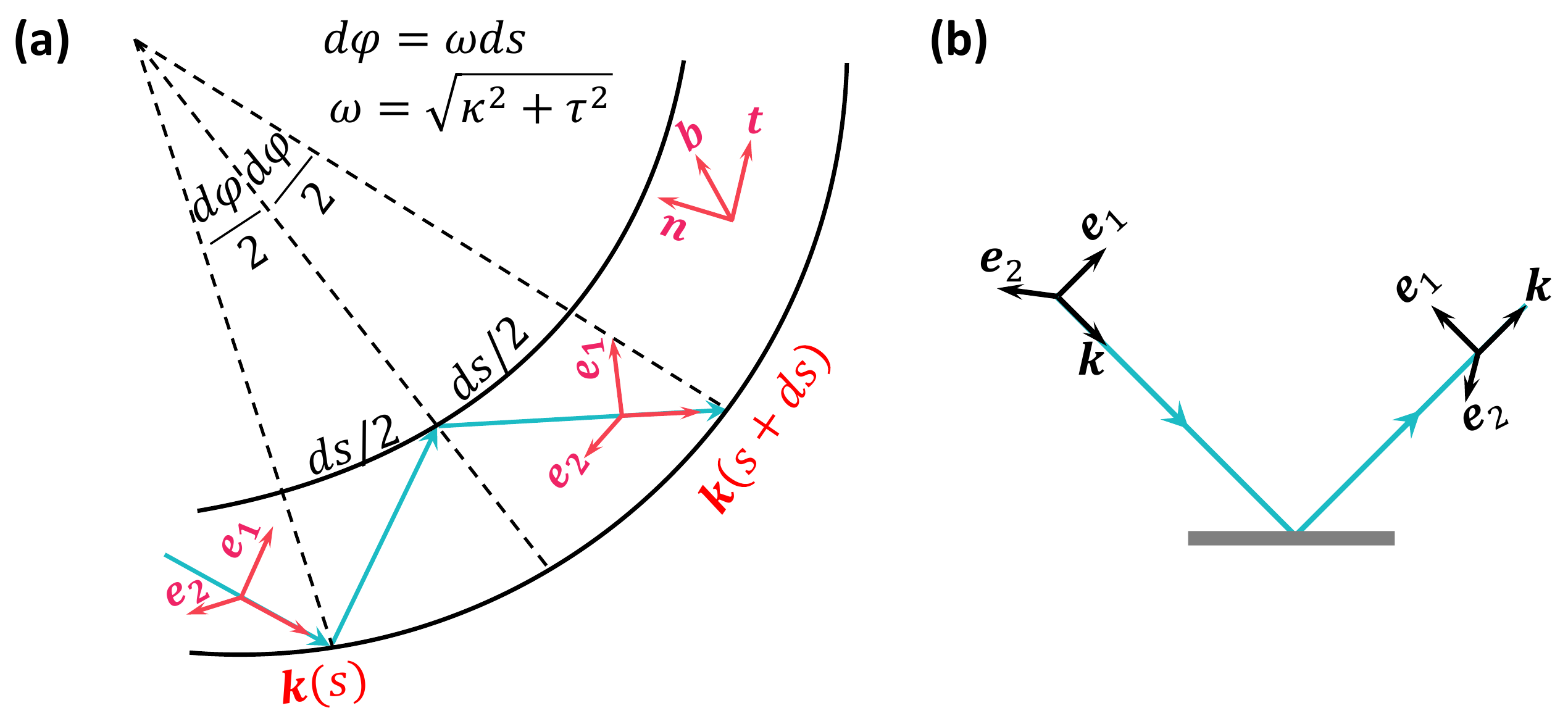}\caption{\label{fig:3} (a) Two successive reflections in the fiber give an adiabatic transformation of the photon coordinate frame (PCF). (b) The change of the PCF under a reflection.}
\end{figure}

\textit{Rotation of the photon coordinate frame}---
The wave vector $\boldsymbol{k}$ and two transverse
polarization unit vectors ($\boldsymbol{e}_{1}$ and $\boldsymbol{e}_{2}$)
form another coordinate frame---the PCF [see Fig.~\ref{fig:3} (a)], which co-moves with the light. We emphasize that the PCF is not synchronized with the LCF usually. This de-synchronization leads to the rotation of the linear polarization vector of light and the geometric phase for circularly polarized light.  We note that it is invalid to evaluate the propagating phase factor via ray optics for single-mode fibers. However, the propagating phase is irrelevant to our concerned problem~\cite{Chiao1986manifestations} and has been ignored. We only focus on the extra geometric phase due to the non-vanishing curvature and torsion of 
a coiled fiber. Here, we assume that the adiabatic change of the wave vector $\boldsymbol{k}$ can be treated as successive reflections in coiled fibers. Previous experimental results~\cite{ross1984rotation,Tomita1986observation} can be perfectly explained with our purely geometric approach. The validity of our assumption can be  further tested by measuring the geometric phase for twisted light as shown in the next section.

Two successive
reflections of light give an adiabatic transformation
of the PCF in the fiber as shown in Fig.~\ref{fig:3} (a). Now we sit on the LCF to study the dynamics of an arbitrary vector $\boldsymbol{v}$ co-moving with the PCF. The reflection of the PCF at $s$ is described by the matrix
\begin{equation}
\hat{P}_{t}=\left[\begin{array}{ccc}
-1 & 0 & 0\\
0 & -1 & 0\\
0 & 0 & 1
\end{array}\right],
\end{equation}
which is the space reversion in the normal plane (i.e., $\boldsymbol{n}\boldsymbol{b}$-plane). Here, we have used the fact that the no rotation of the PCF around the $\boldsymbol{k}$-axis occurs under a reflection as shown in Fig.~\ref{fig:3} (b). From $s$ to $s+ds/2$, the vector $\boldsymbol{v}$ itself remains unchanged. However, the LCF has been rotated by $\hat{R}(\Delta\varphi/2,\boldsymbol{e}(s))$. Thus, from the perspective of the LCF, $\boldsymbol{v}$ has been rotated by $\hat{R}(-\Delta\varphi/2,\boldsymbol{e}(s))$. An infinitesimal adiabatic evolution of the PCF is described by
\begin{equation}
\hat{U}(\Delta\varphi)=\hat{R}\left(-\frac{\Delta\varphi}{2},\boldsymbol{e}(s)\right)\hat{P}_{t}\hat{R}\left(-\frac{\Delta\varphi}{2},\boldsymbol{e}(s)\right)\hat{P}_{t}.\label{eq:U}
\end{equation}
We note that this infinitesimal evolution operator works for all vectors co-moving with the PCF and it is not limited to one specific wave vector or ray path.

If a fiber is wound in a co-planar path (i.e., $\tau (s) =0$), the polarization vector propagates by parallel transport as the same as the LCF. This was taken as an axiom by Ross and supported by his experiment~\cite{ross1984rotation}. Haldane pointed out that this claim follows geometrically by approximating a general fiber path as a sequence of curved co-planar segments joined by straight segments~\cite{haldane1986path}. We now verify this result directly by setting $\theta=\pi/2$ in Eq.~(\ref{eq:U}). We find that the adiabatic evolution operator always equals the identity matrix, i.e., $\hat{U}(\Delta\varphi)=\hat{I}$. Thus, in the absence of torsion, the PCF maintains
synchronization with the LCF. This is different from Chiao and Wu's theory, which implies that the rotation of the polarization vector occurs even in the absence of torsion.

Now we show that the in presence of torsion, the vector $\boldsymbol{v}$ co-moving with the PCF rotates around the $\boldsymbol{t}$-axis during propagation. The motion equation of $\boldsymbol{v}(\varphi)$ in the LCF is given by 
\begin{equation}
\frac{d}{d\varphi}\boldsymbol{v}(\varphi)=\hat{\mathcal{H}}(\varphi)\boldsymbol{v}, \label{eq:EOM_v}
\end{equation}
where
\begin{equation}
\hat{\mathcal{H}}(\varphi)=\lim_{\Delta\varphi\rightarrow0}\frac{\hat{U}(\Delta\varphi)-\hat{I}}{\Delta\varphi}=\left[\begin{array}{ccc}
0 & -\cos\theta(\varphi) & 0\\
\cos\theta(\varphi) & 0 & 0\\
0 & 0 & 0
\end{array}\right].
\end{equation}
The dynamics of the vector $\boldsymbol{v}(\varphi)$ co-moving with light is given by the evolution operator
\begin{equation}
\hat{U}(\varphi)=\left[\begin{array}{ccc}
\cos\alpha(\varphi) & -\sin\alpha(\varphi) & 0\\
\sin\alpha(\varphi) & \cos\alpha(\varphi) & 0\\
0 & 0 & 1
\end{array}\right], \label{eq:U_varphi}
\end{equation}
with $\alpha(\varphi)=\int_0^{\varphi}\cos\theta(\varphi')d\varphi'$. We see that $\hat{U}(\varphi)$ describes the rotation around $\boldsymbol{t}$-axis by angle $\alpha (\varphi)$. The evolution operator can be obtained by re-expressing Eq.~(\ref{eq:EOM_v}) as an effective Schr\"odinger equation. Then, the corresponding evolution operator is given by
$\hat{U}(\varphi)=\hat{M}\hat{\Lambda}(\varphi)\hat{M}^{\dagger}$ with
\begin{equation}
\hat{M}=\frac{1}{\sqrt{2}}\left[\begin{array}{ccc}
1 & 1 & 0\\
i & -i & 0\\
0 & 0 & \sqrt{2}
\end{array}\right],
\end{equation}
formed by the eigenvectors of $i\hat{\mathcal{H}}(\varphi)$ and the diagonal matrix carrying the dynamic phase
\begin{equation}
\hat{\Lambda}(\varphi)=\left[\begin{array}{ccc}
e^{-i\alpha (\varphi)} & 0 & 0\\
0 & e^{i\alpha (\varphi)} & 0\\
0 & 0 & 1
\end{array}\right].
\end{equation}
We note that the infinitesimal adiabatic transformation (\ref{eq:U}) of the PCF includes two reflections, thus no handedness change manifests in Eq.~(\ref{eq:U_varphi}).

For a single-mode
fiber, we assume that light enters the fiber nearly parallel to its tangent vector, i.e., $\boldsymbol{k}(0)/|\boldsymbol{k}|\approx\boldsymbol{t}(0)$. The PCF and LCF coincide at the beginning. It follows that the rotation angle of the polarization vector, i.e., the geometric phases for the left ($+$) and right ($-$) circularly polarized light after an adiabatic circulation are given by
\begin{equation}
\gamma_{\pm}(\mathcal{C})=\pm\alpha (2\pi)=\pm\int_{0}^{2\pi}\cos\theta(\varphi)d\varphi.    
\end{equation} 
In addition to the quantum interpretation from Chiao and Wu~\cite{Chiao1986manifestations}, our method gives a purely geometric explanation of Tomita and Chiao's experiment~\cite{Tomita1986observation}. For a uniform helix, the pitch angle $\theta$ is a constant. Then our phase geometric
phase $\gamma_{\pm}(\mathcal{C})=\pm2\pi\cos\theta$  recovers Chiao and Wu's results~\cite{Chiao1986manifestations}. We emphasize that, in our description, $\gamma_{\pm}(\mathcal{C})$ is due to the purely classical rotation between the two coordinate frames. Even if we treat Eq.~(\ref{eq:EOM_v}) as an effective Schr\"odinger equation, this phase corresponds to the dynamic phase, not the Pancharatnam-Berry phase.

\textit{Geometric phase for twisted light}---The upper analysis is not limited to single-mode fibers. We now apply our method to a twisted light carrying non-zero OAM traveling in a coiled multi-mode fiber. We show that the geometric rotation also occurs in the OAM states. The corresponding geometric phase for twisted light carry $m\hbar$ OAM is given by 
\begin{equation}
\gamma_{m}(\mathcal{C}) =  m \int_0^{2\pi}\cos\theta (\varphi) d\varphi, \label{eq:GPhase}  
\end{equation}
which is $m$ times as large as that of a photon spin state. 

A linearly polarized twisted optical pulse or beam can be generally described by a spectral amplitude function in $\boldsymbol{k}$-space~\cite{Yang2021nonclassical,enderlein2004unified}
\begin{equation}
\xi_{m}(\boldsymbol{k})=\eta(k_{3},\rho_{k})e^{im\phi_{k}}, \label{eq:SAF}
\end{equation}
where $k_3$ denotes the component of $\boldsymbol{k}$ in the propagating direction, $\rho_k=\sqrt{k_1^2+k_2^2}$, $\phi_k$ is the azimuthal angle of $\boldsymbol{k}$ in the PCF, $m$ is an integer determining the OAM quantum number, and the function $\eta$ characterizes the spatial distribution of light pulse. As shown in the previous section, any vector co-moving with the PCF will be rotated around the $\boldsymbol{t}$-axis via $\hat{U}(\varphi)$. Usually, the beam is injected into the fiber along $\boldsymbol{t}$-axis, i.e., $\boldsymbol{e}_3 = \boldsymbol{t}(0)$. We now let the LCF and the PCF coincide with each other at the beginning $\varphi = 0$, i.e., $\boldsymbol{e}_1=\boldsymbol{n}(0)$ and $\boldsymbol{e}_2=\boldsymbol{b}(0)$. Applying the evolution operator $\hat{U}(\varphi)$ on the vector $\boldsymbol{k}$, we have the components of $\boldsymbol{k}$ in the LCF,
\begin{align}
k_n(\varphi) & = k_1\cos\alpha (\varphi)+k_2\sin\alpha (\varphi), \\
k_b(\varphi) & =  -k_1\sin\alpha (\varphi)+k_2\cos\alpha (\varphi), \\
k_t (\varphi) & = k_3.
\end{align}
and the transformation relation
\begin{equation}
(k_1+ik_2)^m\rightarrow [k_n(\varphi)+ik_b(\varphi)]^m\exp[im\alpha (\varphi)]. 
\end{equation} 
The LCP returns to its initial configuration after an adiabatic circulation, i.e., $\boldsymbol{n}(2\pi)=\boldsymbol{e}_1$, $\boldsymbol{b}(2\pi)=\boldsymbol{e}_2$, $\boldsymbol{t}(2\pi)=\boldsymbol{e}_3$. Then, we have \begin{equation}
(k_1+ik_2)^m\xrightarrow[]{\varphi =2\pi} (k_1+ik_2)^m\exp[i\gamma_m (C)]. 
\end{equation}
By re-expressing the phase factor $\exp(im\phi_k)=[(k_1+ik_2)/\rho_k]^m$, we obtain the geometric phase in Eq.~(\ref{eq:GPhase}) for $\xi_m(\boldsymbol{k})$. After a Fourier transformation, this phase factor $\exp[i\gamma_m(\mathcal{C})]$ transfers to the corresponding wave-packet function in the real-space~\cite{yang2022quantum}. We note that our approach works for both a twisted laser beam and a single-photon pulse~\cite{Yang2021nonclassical}.

\begin{figure}
\centering
\includegraphics[width=6cm]{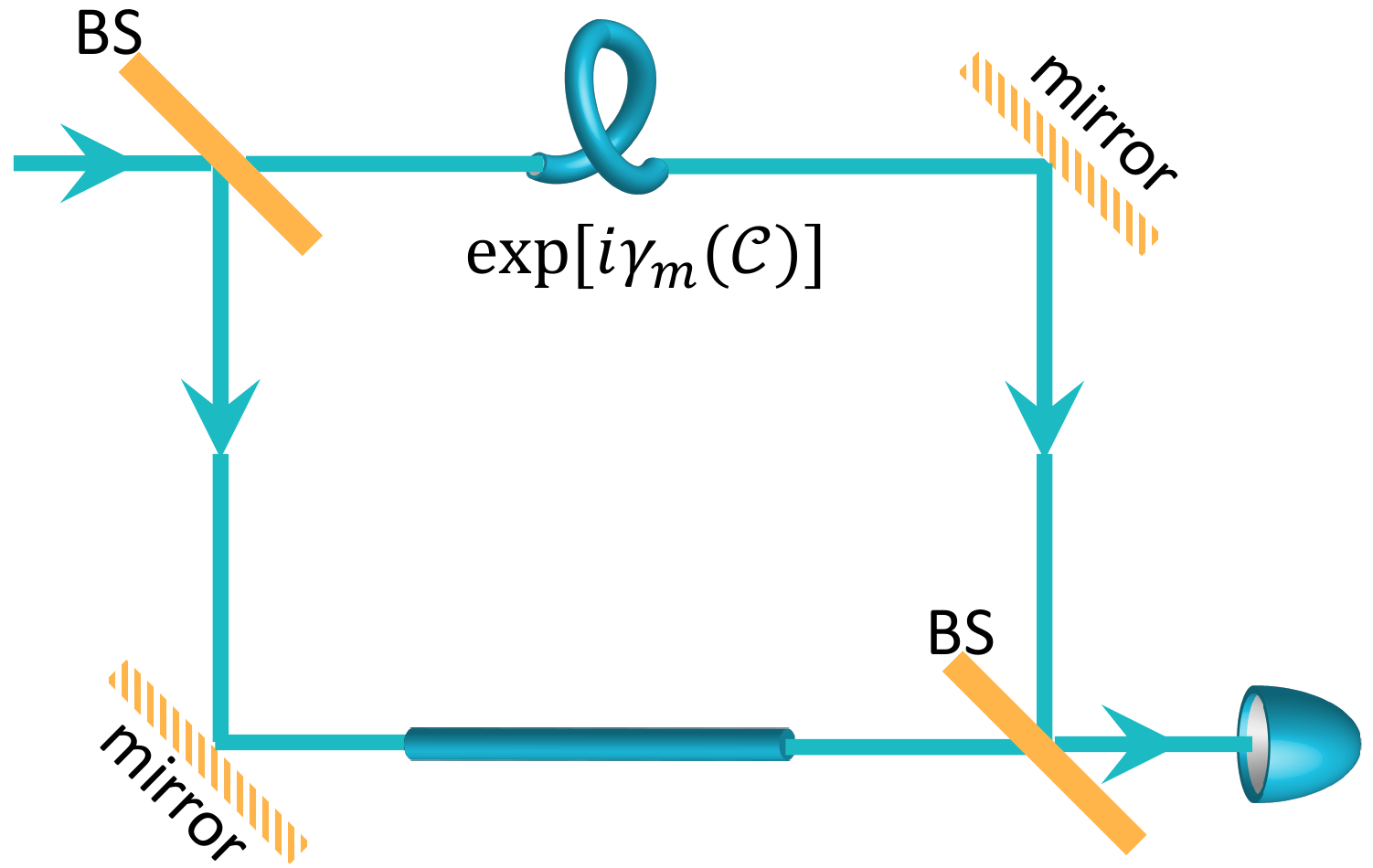}\caption{\label{fig:4} Detection of the geometric phase $\gamma_m (C)$ for twisting light carrying orbital angular momentum via Mach–Zehnder interference. In experiments, an extra Dove prism will be added in the lower optical path after the first beam splitter (BS). }
\end{figure}

Now we shed light on the geometric phase for OAM modes in optical fibers. Under the paraxial approximation, the electric field of a linearly polarized OAM mode in optical fibers can be separated into radial-dependent, azimuthal-dependent, and propagating-phase parts~\cite{alexeyev1998optical,brunet2017optical},
\begin{equation}
\boldsymbol{E} (\rho,\phi,s) = \boldsymbol{\varepsilon}(\rho)\exp(i m\phi)\exp(i\beta s)    
\end{equation} 
where the vector $\boldsymbol{\varepsilon}(\rho)$ characterizes the polarization and radial distribution of the electric field, $m$ is the OAM quantum number, and $\beta$ is the propagation constant~\cite{alexeyev1998optical}. After a Fourier transformation, we have $\boldsymbol{E} (\rho_k,\phi_k,k_3) \propto e^{i m\phi_k}$. The OAM mode function will accumulate a geometric phase $\gamma_m (C)$ after an adiabatic circulation. We note that ray optics is usually not valid to evaluate the propagating phase $\beta s$ of fiber modes. However, the propagating phase is irrelevant to our concerned geometric phase $\gamma_m(\mathcal{C})$, which purely comes from the helical phase factor $\exp (im\phi)$ due to the classical rotation between the LCF and the PCF.

\textit{Measuring geometric phase $\gamma_m(\mathcal{C})$ via optical interference}---The geometric phase in Eq.~(\ref{eq:GPhase}) can be observed in optical interference experiments, such as the Mach-Zehnder interference with an OAM laser beam as shown in Fig.~\ref{fig:4}. There are a straight fiber and a coiled fiber with equal length in the two interference channels, respectively. Our geometric phase $\gamma_m(\mathcal{C})$ will lead to the rotation of the interference petals pattern~\cite{guo2021interference}.

The Mach-Zehner interference of twisted laser beams has been widely studied in experiments~\cite{guo2016measuring,Kumar2019Mach}. We note that in the twisted-light Mach-Zehner interference, an extra Dove prism is employed to reverse the sign of the OAM quantum number in one of the optical paths. Thus, the photon number density measured at the output port is given by
\begin{equation}
\langle \hat{\psi}^{\dagger}(\boldsymbol{r})\hat{\psi}(\boldsymbol{r})\rangle\propto \left|\exp[im\varphi+i\gamma_m(\mathcal{C})]+\exp(-im\varphi)\right|^2,    
\end{equation}
where $\hat{\psi}(\boldsymbol{r})$ is the effective field operator of the paraxial laser beam~\cite{yang2022quantum}. Here, we see that the geometric phase difference $\gamma_m(\mathcal{C})$ between a straight fiber and a coiled fiber leads to the rotation of the interference pattern.
Similar analysis can be applied to a Mach-Zehnder interferometer with entangled twisted photons~\cite{Jha2011Supersensitive}.

\textit{Discussion}---Finally, we note that the geometric phase $\gamma_m(\mathcal{C})$ can be utilized to construct a Z-gate for OAM-state-based quantum computation~\cite{babazadeh2017high}, which has been previously realized via a pair of Dove prisms~\cite{de2005implementing,wang2015Quantum,Zhang2016Engineering}. Previously, entangled twisted photons have been exploited for the super-sensitive measurement of angular displacements~\cite{Jha2011Supersensitive}. The geometric phase for twisted photons can be used to detect the pitch-angle-related quantities with higher sensitivity via a similar interference strategy~\cite{Maga2014amplification}.

\textit{Acknowledgments.} The author thanks Professor C. P. Sun for bringing my attention to this issue and S. W. Li for the fruitful discussion. This work is funded by National Key R\&D Program of China (Grant No. 2021YFE0193500) and NSFC Grant
No.12275048.

\bibliography{main}
\end{document}